# Unusual Ferroelectricity Induced by the Jahn-Teller Effect: A Case Study on Lacunar Spinel Compounds

Ke Xu[1,2] and H. J. Xiang[1*]

[1]Key Laboratory of Computational Physical Sciences (Ministry of Education), State Key Laboratory of Surface Physics, Collaborative Innovation Center of Advanced Microstructures, and Department of Physics, Fudan University, Shanghai 200433, P. R. China

[2]Hubei Key Laboratory of Low Dimensional Optoelectronic Materials and Devices, Hubei University of Arts and Science, Xiangyang, 441053, P. R. China

Email: hxiang@fudan.edu.cn

The Jahn-Teller effect refers to the symmetry-lowering geometrical distortion in a crystal (or non-linear molecule) due to the presence of a degenerate electronic state. Usually, the Jahn-Teller distortion is not polar. Recently, $GaV_4S_8$ with the lacunar spinel structure was found to undergoes a Jahn-Teller distortion from cubic to ferroelectric rhombohedral structure at $T_{JT} = 38K$. In this work, we carry out a general group theory analysis to show how and when the Jahn-Teller effect gives rise to ferroelectricity. On the basis of this theory, we find that the ferroelectric Jahn-Teller distort in $GaV_4S_8$ is due to the non-centrosymmetric nature of the parent phase and strong electron-phonon interaction related to two low-energy $T_2$ phonon modes. Interestingly, $GaV_4S_8$ is not only ferroelectric, but also ferromagnetic with the magnetic easy axis along the ferroelectric direction. This suggests that $GaV_4S_8$ is a multiferroic in which an external electric field may control its magnetization direction. Our study not only explains the Jahn-Teller physics in $GaV_4S_8$, but also paves a new way for searching and designing new ferroelectrics and multiferroics.

In 1937, Jahn and Teller performed the symmetry analysis to demonstrate that the nonlinear polyatomic molecule with degenerated orbitals is not stable and it will spontaneously distort itself in some way to remove its degeneracy [1]. Later, the Jahn-Teller (JT) effect was found to also take place in solids as a special case of the electron-phonon coupling [2]. It turns out that the JT effect is relevant to many exotic phenomena in condensed matter physics. For example, some models have been put forward to understand the role of the JT effect in the high-Tc superconductivity in cuprates [3-9]. The JT effect is also a key ingredient for understanding the colossal magnetoresistance in manganites [10-13].

Usually, the JT distortion is of the non-polar nature, i.e., it does not induce the ferroelectricity. Perovskite $KCuF_3$ and $LaMnO_3$ are two typical examples in which the cooperative JT effect takes place and induce the orbital ordering [12-14]. At high temperature, both $KCuF_3$ and $LaMnO_3$ are cubic with the Pm-3m space group. The low temperature phases of $KCuF_3$ and $LaMnO_3$ take I4/mcm [14,15] and Pnma [10] space groups, respectively. Although their symmetry is lowered by the JT distortion, the centrosymmetry remains intact. The above experimental fact makes most people believe that the JT effect could not give rise to proper ferroelectricity. Note that here we focus on the original JT effect, instead of the pseudo-JT effect (or second-order JT effect), which is responsible for the ferroelectricity in many common ferroelectrics such as $BaTiO_3$ [17].

Interestingly, a recent study indicates that the JT distortion in $GaV_4S_8$ leads to a ferroelectric polarization (~0.6 $\mu C \cdot cm^{-2}$) along the [111] direction [18,19]. $GaV_4S_8$ belongs to the family of lacunar spinels $AM_4Q_8$ compounds [20-23] (A = Ga and Ge; M = V, Mo, Nb and Ta; X = S and Se), which attract enormous research attention since they exhibit abundant physical phenomena, such as high pressure induced superconducting phase transition [24] and correlation-driven insulator-to-metal transition [25,26]. At room temperature, $GaV_4S_8$ takes the high symmetry lacunar spinel structure with the $T_d$ symmetry (i.e., non-centrosymmetric and non-polar). At 42 K, the cooperative JT distortion drives a cubic to rhombohedral ($C_{3v}$ symmetry) structural transition. We note that a similar JT distortion induced ferroelectricity was also observed in the lacunar spinel compound $GeV_4S_8$ [27]. Up to now, the understanding of the JT distortion induced ferroelectricity in lacunar spinel compounds is incomplete. The answer to this question is not only relevant to the lacunar spinel compounds, but also may lead to new direction for designing/predicting new ferroelectrics and multiferroics.

In this rapid communication, we explore the possibility of inducing ferroelectricity through the JT effect. A general group theory analysis is performed to show that the JT effect might induce the ferroelectricity if the parent structure belongs to one of the 10 non-polar non-centrosymmetric point groups. By combining group theory analysis and first-principles calculations on the electron-phonon coupling constants, we find that the non-centrosymmetric nature of cubic $GaV_4S_8$ and the presence of two low-energy $T_2$ modes with strong electron-phonon coupling are crucial to the occurrence of ferroelectricity induced by the JT distortion. We propose that the magnetic easy axis can be controlled by the external magnetic field in multiferroic R3m $GaV_4S_8$. Finally, we predict that the JT effect will also induce ferroelectricity in $GaMo_4S_8$ and R3m $GaMo_4S_8$ is a ferromagnetic multiferroic.

Let's assume that a parent high symmetry structure ($G_0$ space group) with the equilibrium configuration $Q_0$ has a degenerate electronic ground state, i.e, a n-fold degenerate level of the original electronic Hamiltonian $H_0$ is partially filled. The single particle wavefunctions of the degenerate states are denoted by $\phi_i$ (i = 1, n). The JT effect results in a lower symmetry nuclear configuration Q, which can obtain from the high symmetry equilibrium configuration $Q_0$ by adding the linear combination of phonon normal modes: $Q = Q_0 + \sum_r \eta_r Q_r$ ($\eta_r$ is the magnitude of the normal mode $Q_r$). The electronic Hamiltonian of the distorted structure can be expanded as a power series in terms of $\eta_r$. For our current purpose, it is adequate to consider the terms up to the linear order:

$$H = H_0 + H' = H_0 + \sum_r \frac{\partial V}{\partial \eta_r} \eta_r, \quad (1)$$

where $V$ is the one-electron crystalline potential. To investigate how the distortion splits the degenerate level, one can adopt the degenerate perturbation theory by diagonalizing the $n \times n$ matrix with matrix elements defined below:

$$H'_{ij} = \sum_r \eta_r \left\langle \phi_j \left| \frac{\partial V}{\partial \eta_r} \right| \phi_i \right\rangle, \quad (2)$$

For simplicity, we consider the $\Gamma$ normal modes and $\Gamma$ electronic states. In this case, the point group $K_0$ of the space group $G_0$ can be used to classify $\phi_i$ and $\frac{\partial V}{\partial \eta_r}$. Note that $\frac{\partial V}{\partial \eta_r}$ has the same symmetry transformation property as the normal mode $Q_r$. We assume that $\phi_i$ and $Q_r$ transform as the irreducible representations (IRs) $\hat{D}_e$ and $\hat{D}_p$, respectively. According to the group theory, only when the product $\phi_j^* \frac{\partial V}{\partial \eta_r} \phi_i$ contains the identical representation, or equivalently, the symmetrical product ($\hat{D}_e^{\{2\}}$) of $\hat{D}_e$ contains $\hat{D}_p$, the integral $\left\langle \phi_j \left| \frac{\partial V}{\partial \eta_r} \right| \phi_i \right\rangle$ can be non-zero. Otherwise, the degenerate electronic states will not split, and the JT distortion will not take place.

If the point group $K_0$ of the parent structure contains the spatial inversion symmetry, one can easily show that the JT effect will not break the centrosymmetry. For a point group containing the inversion symmetry, the IR is either even (g) or odd (u). The IR for $\phi_i$ can be even or odd. In any cases, the symmetrical product $\hat{D}_e^{\{2\}}$ only contains even IRs. If the normal mode $Q_r$ has the odd symmetry, $\left\langle \phi_j \left| \frac{\partial V}{\partial \eta_r} \right| \phi_i \right\rangle$ will be zero. This means that the JT distortion should not contain odd normal modes when the parent structure is centrosymmetric. Thus, we explain why the JT distortion cannot induce ferroelectricity in the centrosymmetric systems such as $KCuF_3$ and $LaMnO_3$. The JT distortion was found to induce local electric polarization in $DyVO_4$ and $PrCl_3$ because the site-symmetry of $Dy^{3+}$ and $Pr^{3+}$ is non-centrosymmetric. However, the overall polarization cancels, resulting in an antiferroelectric structure

[28,29]. This is because the point group of the high temperature structures of DyVO$_4$ and PrCl$_3$ contains the centrosymmetry. Our group theory analysis shows that the overall symmetry of the parent structure instead of the local site-symmetry should be adopted to predict whether the JT effect can induce ferroelectricity.

As we discussed above, the JT effect can not induce ferroelectricity if the parent structure is centrosymmetric. If the parent structure itself is polar, the JT effect may give rise to an additional electric polarization. However, the electric polarization may not be switched. Below, we will focus on the case where the parent structure is non-polar and non-centrosymmetric. Of the 32 crystal point groups, there are 11 non-polar and non-centrosymmetric point groups, namely, D$_2$, S$_4$, D$_4$, D$_{2d}$, D$_3$, C$_{3h}$, D$_6$, D$_{3h}$, T, O, and T$_d$ point group. Since the D$_2$ point group does not have any degenerate IRs, there is no JT effect in a parent structure with such a point group symmetry. For each of the other 10 point groups, we consider each degenerate IR $\hat{D}_e$ for the electronic states. Then the symmetrical product $\hat{D}_e^{\{2\}}$ is decomposed into IRs from which the IR $\hat{D}_p$ of the normal mode of the JT distortion is selected. If general, the normal mode of a given IR $\hat{D}_p$ is characterized by its order parameter direction. For each possible normal mode of the IR $\hat{D}_p$, we find out the point group of the final distorted structure to tell whether the JT effect induces ferroelectricity. All these results are listed in [30]. In particular, we find that the JT distortion will always induce the ferroelectricity when the parent structure with S$_4$, D$_3$, C$_{3h}$, or D$_{3h}$ point groups has a degenerate electronic state and the cell size is not enlarged. Therefore, the JT effect may induce ferroelectricity when the point group of the parent structure is non-polar and non-centrosymmetric. This is one of our main results.

Now we begin to address why the JT effect gives rise to ferroelectricity in GaV$_4$S$_8$. At room temperature, cubic GaV$_4$S$_8$ exhibits F-43m space group and the four V atoms occupy the 16e Wyckoff position with the site symmetry C$_{3v}$, which form a regular tetrahedral structure (see Fig. 1a) and the V-V bond length is about 2.89 Å (Fig. 1c). The band structure and schematic illustration of the energy levels of cubic

GaV$_4$S$_8$ are shown in Fig. 2. Here, the spin unpolarized state is considered. Our calculations show that there is a narrow partially filled three-fold degenerate t$_2$ band. The fat band analysis shows that this degenerate band is mainly contributed from the 3d orbitals of the V$_4$ cluster. In GaV$_4$S$_8$, each V ion is in +3.25 valence state. Thus, there are seven valence electrons per V$_4$ cluster, among which six electrons fully occupy the low-energy a$_1$ and e levels (see Fig. 2b) and the remaining electron partially fills the three-fold degenerate t$_2$ level [31]. This indicates that GaV$_4$S$_8$ is a JT active system.

Since cubic GaV$_4$S$_8$ with the non-centrosymmetric T$_d$ point group has a degenerate electronic state on the t$_2$ level, the JT effect may give rise to ferroelectricity according to our above group theory analysis. The symmetrical product of t$_2$ can be decomposed as: $t_2^{\{2\}} = a_1 \oplus e \oplus t_2$. Therefore, the active normal mode can possibly belong to A$_1$ ($\Gamma_1$), E ($\Gamma_3$), or T$_2$ ($\Gamma_4$). Since A$_1$ is a fully symmetric representation which cannot split the electronic level, this mode is not relevant. We find that the T$_2$ mode can induce the ferroelectricity, while the E mode cannot (see Fig. 3). The group theory can only show that there is a possibility to induce the ferroelectricity by the JT effect in GaV$_4$S$_8$ since it can not determine which normal mode (E or T$_2$) will be active.

In order to determine the exact normal mode in the low temperature GaV$_4$S$_8$ structure, we will minimize the total energy with respect to the distortion. Within our simple model, the total energy of the distorted structure contains two parts:

$$E(Q) = E_{strain}(Q) + E_{JT}(Q), \quad (3)$$

where $E_{strain}(Q)$ is the elastic strain energy cost and $E_{JT}(Q)$ is the electronic energy gain due to the JT effect. Using the eigenvectors of the force constant matrix (not dynamic matrix) as the normal modes $Q_r$, the strain energy can be written as $E_{strain} = \sum_r \omega_r \eta_r^2$, where $\omega_r$ is the eigenvalue of the force constant matrix. The electronic energy $E_{JT}(Q)$ is the sum of eigenvalues of the occupied states:

$E_{JT}(Q) = \sum_{i=1}^{n} f_i \varepsilon_i(Q)$, where $f_i$ is the Fermi-Dirac distribution and $\varepsilon_i$ is the eigenvalue of the electronic states for the distorted nuclear configuration $Q$. To obtain $\varepsilon_i$, we diagonalize the perturbed Hamiltonian $H'$ with the matrix elements defined in Eq. (2). In practical calculations, the electron-phonon matrix element $g_{ij}^r = \langle \phi_j | \frac{\partial V}{\partial \eta_r} | \phi_i \rangle$ is computed through the density functional perturbation theory. We minimize the total energy with respect to $Q$ (in fact, the magnitude of the normal modes $\eta_r$) by combining the Monte-Carlo annealing with conjugate gradient local optimization (see [30]).

Each $GaV_4S_8$ primitive cell contains one formula unit. So there are 39 phonon normal modes, which can be classified into 3 $T_1$ modes, 7 $T_2$ modes, 3 E modes, and 3 $A_1$ modes (Fig. 4a). In cubic $GaV_4S_8$, the partially filled degenerate electronic $t_2$ state is three-fold, i.e., n = 3. As already mentioned, we consider only $\Gamma$ normal modes and $\Gamma$ electronic states. The magnitude of the electron-phonon matrix element is shown in Fig. 4b. We find that two low frequency $T_2$ normal modes (at around 0.2 and 0.3 Ry/Bohr$^2$) couple strongly to the electronic $t_2$ state. Note that the lower $T_2$ mode has the lowest frequency except for the zero-frequency acoustic modes. The lowest frequency $E_2$ mode (at 0.33 Ry/Bohr$^2$) couples weaker to the electronic $t_2$ state than the nearby $T_2$ mode. The electron-phonon matrix elements for the $T_1$ modes (not shown) are found to be negligible, in agreement with the group theory analysis.

Through minimizing the total energy $E(Q)$, we obtain the distorted structure of $GaV_4S_8$. The contribution to the distortion from each normal mode is shown in Fig. 4c. It can be seen that the two low-energy $T_2$ modes are dominant with some tiny contribution from other $T_2$ modes and $A_1$ modes. Due to the JT effect, the system distorts according to the $T_2$ modes with the order parameter direction along (a,a,a), lowering the symmetry of $GaV_4S_8$ from $T_d$ to $C_{3v}$ and resulting in ferroelectricity, in agreement with the experimental results. In the meanwhile, the original degenerate $t_2$ electronic state splits into an occupied nondegenerate $a_1$ level and an unoccupied two-fold degenerate e level, as shown in Fig. 2b. It is interesting to find that there is no contribution to the distortion from the E modes, despite the fact that the electron-phonon coupling related to the E modes is substantial. We perform a test calculation to find that a stronger electron-phonon coupling (e.g., two times stronger) for the E mode results in a different distorted structure with the non-polar $D_{2d}$ symmetry. It is clear that there is a competition between the two low-energy $T_2$ modes and E modes in lowering the electronic energy $E_{JT}(Q)$. Because the two $T_2$ modes are softer and the electron-phonon coupling for the $T_2$ modes is even stronger than the E modes, the JT effect induces ferroelectricity in $GaV_4S_8$.

The way that $GaV_4S_8$ distorts can be reasoned in terms of the orbital interaction picture. Fig. 1b shows the $\Gamma$ wavefunction ($a_1$ symmetry) of the highest occupied valence band of the experimental low symmetry R3m $GaV_4S_8$ structure. The $a_1$ state is mainly contributed by the $d_{z2}$ orbitals (here, the local z axis is along the ferroelectric [111] direction) of the four V atoms. We can see that the three $d_{z2}$ orbitals of the lateral V atoms form a three-center bonding, while the $d_{z2}$ orbital of the apex V atom interacts with the other $d_{z2}$ orbitals in an anti-bonding way. To make the occupied $a_1$ state more stable, the three lateral V atoms should move towards to each other, while the apex V atom should move away from the V triangle. That is exactly what is found experimentally (see Fig. 1c). In the low temperature R3m structure, the distance between the lateral V atoms is 2.85 Å, which is 0.07 Å shorter than that between the apex V atom and the lateral V atoms (Fig. 1c).

We now turn to the physical properties of the low temperature rhombohedral phase of $GaV_4S_8$. Experiments showed that R3m $GaV_4S_8$ is a ferromagnetic semiconductor. Our DFT+U calculations confirms that it is semiconducting with an indirect band gap of 0.483 eV [30]. The total magnetic moment for the FM state is calculated to be 1 $\mu_B$/f.u., in agreement with the experimental results [32]. With the Berry phase approach, we find that the electric polarization along the [111] direction is 2.43 $\mu C/cm^2$, which is larger than the experimental measured value (about 0.6

μC/cm$^2$). This is reasonable since the band gap of R3m GaV$_4$S$_8$ is not large, which may lead to the charge leakage in the measurement. Since R3m GaV$_4$S$_8$ is simultaneously ferromagnetic and ferroelectric, it is a multiferroic, as recently discovered experimentally. We propose that this multiferroic displays an interesting magnetoelectric coupling. Including the spin-orbit coupling effect, our calculations show that R3m GaV$_4$S$_8$ displays a magnetic anisotropy with the magnetic easy axis along the ferroelectric [111] direction and the anisotropy energy as large as 5.47 meV. The easy axis behavior is in agreement with the experimental results [18]. This indicates that direction of magnetization is always interlocked with the direction of electric polarization. In a cubic system, there are eight [111] directions. This means that there may exist eight different FE domains in the low temperature sample of GaV$_4$S$_8$. If an electric field induce the 109° or 71° switch of the FE polarization, the magnetization easy-axis will also rotate 109° or 71° [30]. This mechanism for the electric field control of magnetism is similar to that proposed in the predicted room temperature multiferroic Zn$_2$FeOsO$_6$ [33]. At last, we predict that GaMo$_4$S$_8$ is also a multiferroic with the similar properties as the GaV$_4$S$_8$ [30].

Note: After the submission of our work, we notice that Barone *et al.* recently discussed the possibility of designing multiferroics based on the JT effect [34]. The proposed materials (Ba$_2$VGe$_2$O$_7$ and Ba$_2$NiGe$_2$O$_7$) were found to have an antiferroelectric ground state.

**Acknowledgements**

Work was supported by NSFC, FANEDD, NCET-10-0351, Research Program of Shanghai Municipality and MOE, the Special Funds for Major State Basic Research, Program for Professor of Special Appointment (Eastern Scholar), and Fok Ying Tung Education Foundation. We thank Prof. Jianjun Xu for the helpful discussion. K. X. was partially supported by NSFC 11404109.

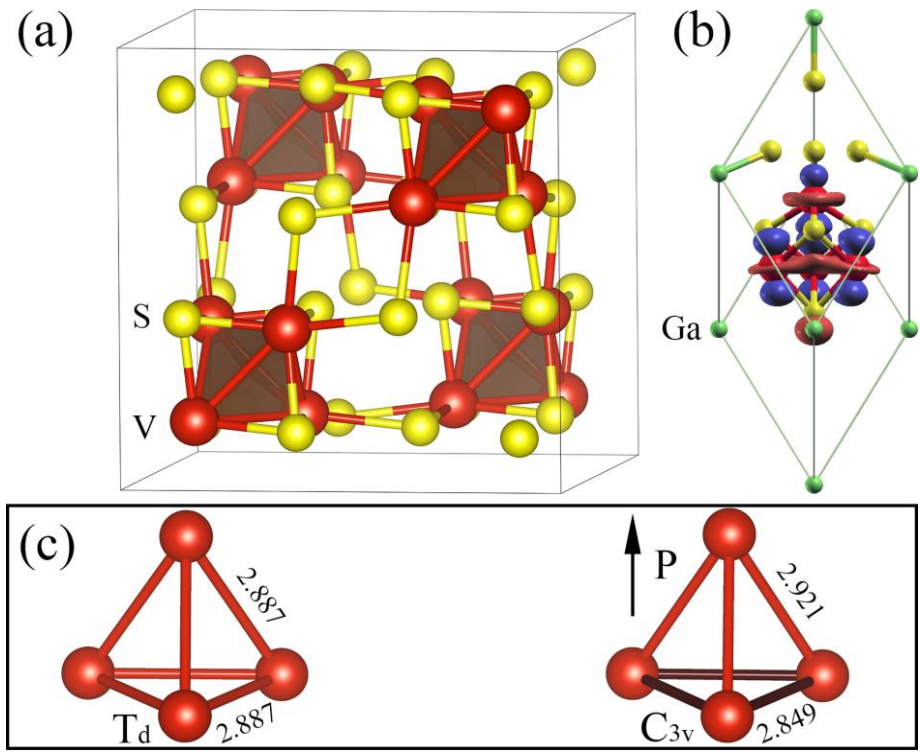

Fig. 1 **Structure and wavefunction of lacunar spinel GaV$_4$S$_8$.** (a) The conventional cell of GaV$_4$S$_8$ with the cubic F-43m symmetry. The V$_4$ tetrahedrons are displayed. Ga atoms are not shown for clarity. (b) The isosurface plot of the highest occupied Γ wavefunction with the a$_1$ symmetry of the low symmetry R3m structure. (c) The comparison of the V$_4$ tetrahedron between cubic T$_d$ GaV$_4$S$_8$ and rhombohedral C$_{3v}$ GaV$_4$S$_8$. The direction of ferroelectric polarization is also shown in the case of C$_{3v}$ GaV$_4$S$_8$.

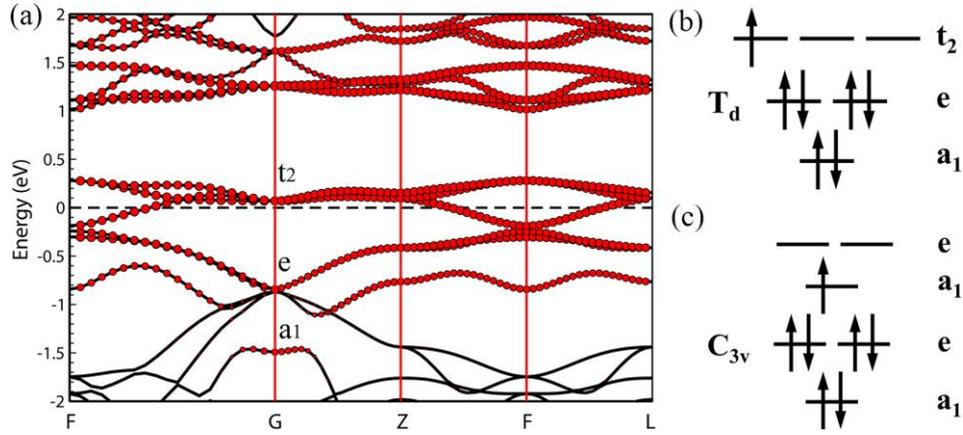

Fig.2 **Electronic properties of $GaV_4S_8$.** (a) The band structure of cubic $GaV_4S_8$ from the spin unpolarized DFT calculation. The Fermi energy is set to zero. Red cycles denote the contributions from the 3d orbitals of V atoms. (b) and (c): Schematic illustrations of the energy levels related to $V_4$ clusters in $T_d$ $GaV_4S_8$ and $C_{3v}$ $GaV_4S_8$, respectively.

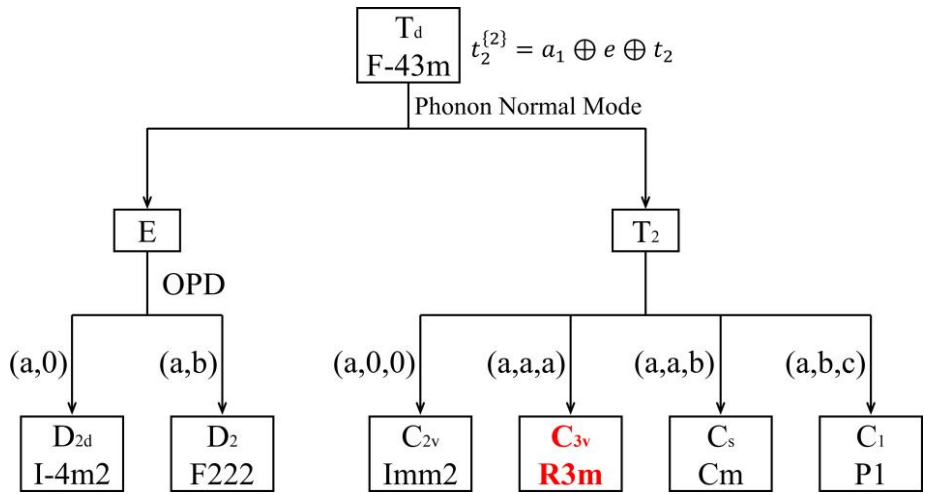

Fig. 3 **Possible JT distortions in a $T_d$ compound with a partially occupied $t_2$ electronic level.** From the symmetrical product of $t_2$, one obtains the JT active irreducible representations (IRs) of phonon normal modes. For each possible phonon IR, the JT active normal mode is characterized by the order parameter direction (OPD). For each normal mode, the symmetry of the resulting distorted structure is denoted. Lacunar spinel $GaV_4S_8$ takes the ferroelectric $C_{3v}$ R3m structure after the JT distortion.

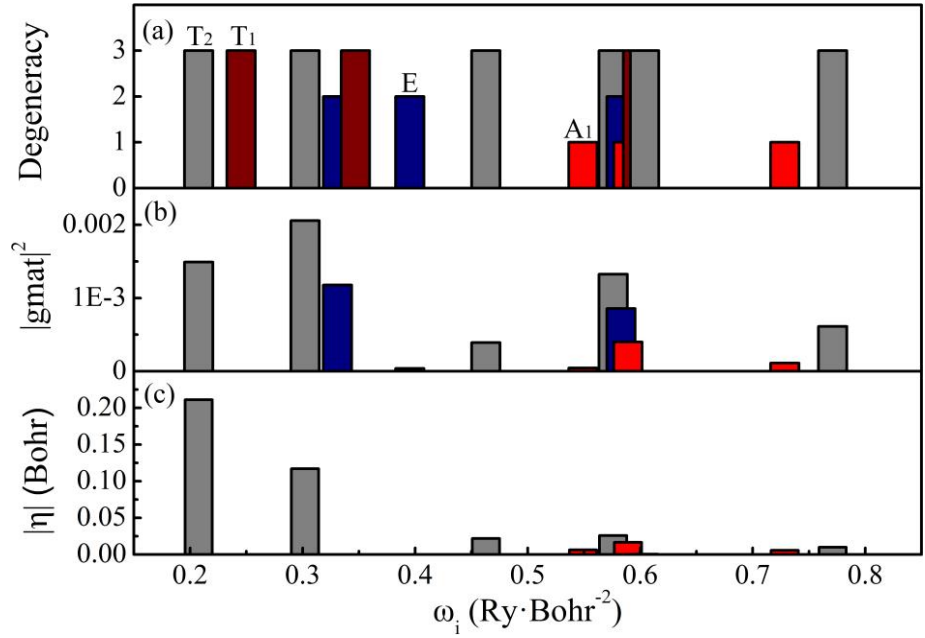

Fig. 4 **Predicted JT distortion in GaV$_4$S$_8$ on the basis of the model including electron-phonon coupling.** (a) The frequencies ($\omega_i$ is the eigenvalue of the force constant matrix) and the degeneracy of normal modes at Γ. The trivial translational modes are not shown. (b) The amplitude of electron-phonon coupling coefficients, which is defined as: $|gmat|^2 = \sum_{i,j,k}|g_{ij}^k|^2$, $g_{ij}^k$ is the electron-phonon matrix element, k is the index for the summation over the dimension of the IR of the phonon normal mode. (c) The magnitude of phonon normal modes in the JT distortion. Most of the phonon normal modes have the T$_2$ symmetry. Gray: T$_2$ mode; wine: T$_1$ mode; navy: E mode; red: A$_1$ mode.